\documentclass{article}
\usepackage{spconfa4}
\usepackage{amsmath,graphicx,hyperref,amsfonts, booktabs,multirow,bm,cleveref,siunitx}
\usepackage{makecell}
\usepackage[export]{adjustbox}
\usepackage{cite}
\usepackage{hyperref}
\usepackage{algorithm}
\usepackage{algpseudocode}

\title{SF-Flow: Sound field magnitude estimation via flow matching\\guided by sparse measurements }

\name{Ege Erdem$^{1, 2}$ 
\sthanks{* Work performed in part during Ege Erdem's internship at NII.} 
\quad Shoichi Koyama$^{2}$ \quad Tomohiko Nakamura$^{3}$  \quad Orchisama Das$^{1}$ \quad Zoran Cvetkovi\'c$^{1}$}
  
  \address{$^{1}$ King's College London, London, UK \quad
      $^{2}$National Institute of Informatics, Tokyo, Japan\\
    $^{3}$National Institute of Advanced Industrial Science and Technology, Tokyo, Japan
    \thanks{This work is supported by JST FOREST Prog. Grant No.\ JPMJFR216M.}
    }

\begin{document}
\ninept
\maketitle

\begin{abstract}
Reconstructing a 3D sound field from sparse microphone measurements is a fundamental yet ill-posed problem, which we address through Acoustic Transfer Function (ATF) magnitude estimation. ATF magnitude encapsulates key perceptual and acoustic properties of a physical space with applications in room characterization and correction.
Although recent generative paradigms such as Flow Matching (FM) have achieved state-of-the-art performance in speech and music generation, their potential in spatial audio remains underexplored. We propose a novel framework for 3D ATF magnitude reconstruction as a guided generation task, with a 3D U-Net conditioned by a permutation-invariant set encoder. This architecture enables reconstruction from an arbitrary number of sparse inputs while leveraging the stable and efficient training properties of FM. Experimental results demonstrate that SF-Flow achieves accurate reconstruction up to \SI{1}{kHz}, trains substantially faster than the autoencoder baseline, and improves significantly with dataset size.
\end{abstract}


\begin{keywords}
sound field reconstruction, flow matching, acoustic transfer function, spatial audio, generative model
\end{keywords}

\vspace{-6pt}
\section{Introduction}
\label{sec:intro}
\vspace{-6pt}

Reconstruction of acoustic fields from sparse measurements is a central challenge in spatial audio. This is typically done in the domain of Room Impulse Responses (RIRs) or Acoustic Transfer Functions (ATFs), which are acoustical fingerprints of an environment and are essential for applications ranging from room acoustics analysis to immersive audio rendering in AR/VR~\cite{Betlehem:IEEE_M_SP2015,Ueno:FnT_SP2025}.
Although the ATF is inherently complex-valued, its magnitude alone captures perceptually and functionally relevant properties of the sound field, including modal resonances, spectral coloration, and frequency-dependent energy decay~\cite{Kuttruff:RoomAcoust}. 
Consequently, magnitude-only reconstruction is valuable when phase estimation is unreliable ({\sl e.g.} with unsynchronized sensors) or of secondary importance, with applications in room characterization and design --- where the spatial distribution of ATF magnitudes captures modal resonances --- such as studio equalization, subwoofer placement, and room-correction systems~\cite{cecchi, welti}, as well as source placement optimization, speech dereverberation, and data augmentation for audio generation systems~\cite{Lluis:JASA2020,Liang:EURASIP2024,miotelloicassp24,Koyama:ForumAcusticum2025,genda_lin25}.

Earlier works focus on estimating ATF distributions based on the principles of underlying physics~\cite{Williams:FourierAcoust}. The sound field is approximated using projections onto spatial basis functions, {\sl e.g.} plane wave functions or spherical wave functions ~\cite{ege19,Colton:InvAcoust_2013,Ueno:FnT_SP2025, das2021room},  
by determining the expansion coefficients as least-squares solutions under the constraints imposed by the Helmholtz equation.
Basis-expansion methods have further evolved into approaches that rely on sparse representations~\cite{ege23, Koyama2019Sparse} and kernel regression~\cite{Ueno:IEEE_J_SP2021}.
Since such conventional methods suffer significant performance degradation when measurements are sparse, deep learning approaches have attracted attention in recent years~\cite{Koyama:IEEE_M_SP2025}. These  can be broadly classified into feedforward neural networks, including autoencoders~\cite{Lluis:JASA2020,Koyama:ForumAcusticum2025}, implicit neural representations, including Physics-Informed Neural Networks (PINN)~\cite{lnaf, Ribeiro:IEEE_ACM_J_ASLP2024, Olivieri:EURASIP2024}, and generative models~\cite{efren2023gan,  antonFastRIR22, miotelloicassp24, diffusionrir}. 
Generative models proved effective in several inverse problems~\cite{elio23}; for the sound field estimation problem, methods based on Generative Adversarial Networks (GANs)~\cite{efren2023gan,  antonFastRIR22} and Denoising Diffusion Probabilistic Models (DDPMs)~\cite{miotelloicassp24, diffusionrir, lin2026gencho} were proposed. 
In \cite{miotelloicassp24}, the authors adapt DDPM-based image inpainting to ATF magnitude reconstruction,
but the method is restricted to 2D distributions and requires relatively dense arrays ($\ge$ 64 microphones). For 3D ATF magnitude estimation, the autoencoder proposed in \cite{Koyama:ForumAcusticum2025}, conditioned on source, microphone positions and frequency, aggregates latent variables into position-independent prototypes.                     

Flow Matching (FM)~\cite{lipman2023flow, liu_rectFM22} has recently emerged as a powerful alternative to diffusion models.
In \cite{brunetto2026flac},  FM is proposed for few-shot RIR synthesis, conditioned on multimodal scene context, including depth maps and acoustic observations, while in \cite{lee2026solving}, a classical analytic Wiener filter is formulated within the FM framework to solve RIR restoration tasks such as denoising and deconvolution. In
\cite{Arellano2025:WASPAA2025} and \cite{promptreverb}, FM is applied to RIR synthesis conditioned on high level acoustic and linguistic descriptors respectively, addressing creative needs rather than physical sound field reconstruction.
In this paper, we introduce \textit{SF-Flow}, a 3D sound field magnitude estimation method built on FM. 
FM offers several advantages for this task: simulation-free training, few-step inference relative to diffusion models, stable training dynamics, and the flexibility to use optimized probability paths, making it a compelling candidate for modeling high-dimensional sound fields. 
Our key contributions are as follows:
(1) we frame 3D sound field magnitude reconstruction as a conditional (guided) generation problem solved with FM, achieving comparable or superior reconstruction with substantially faster training;
(2) we introduce a permutation-invariant set encoder that conditions the generation on an arbitrary number and configuration of sparse microphone measurements;
(3) we provide insights into the impact of dataset scaling in this data-limited application domain.

\vspace{-6pt}
\section{Proposed Method}
\vspace{-6pt}

\subsection{Problem Statement}
\vspace{-4pt}

The ATF, denoted by $H(\mathbf{p}_{\text{src}}, \mathbf{p}_{\text{mic}}, f)$, is a complex-valued function that describes the acoustic response between a source position $\mathbf{p}_{\text{src}}$ and a microphone position $\mathbf{p}_{\text{mic}}$ at a specific frequency $f$. 
Its magnitude, $|H(\cdot)|$, expressed here in decibels (dB), captures modal resonances, spectral coloration, and          
frequency-dependent energy distribution across space.
The problem we address is ATF magnitude estimation within a 3D target region from a limited number of measurements.
We denote the target discretized ATF magnitude as a tensor $\mathbf{H} \in \mathbb{R}^{F \times D \times H \times W}$, where $F$ is the number of frequency bins and $D \times H \times W$ are the depth, height and width of the spatial grid for a fixed source position. The available data are given as a sparse, variable and unordered set of $M$ observations at a given source position,
$\mathcal{C} = \{(\mathbf{g}_i, \mathbf{m}_i)\}_{i=1}^{M}$, where each entry contains
a 9-dimensional geometric descriptor
$\mathbf{g}_i = [\mathbf{r}_i^\top, \mathbf{d}^\top]^\top \in \mathbb{R}^9$,
where $\mathbf{r}_i \in \mathbb{R}^3$ is the  source-to-microphone vector and $\mathbf{d} \in \mathbb{R}^6$ are the source-to-wall distances for a cuboid room,       
and its corresponding $F$-dimensional ATF magnitude vector $\mathbf{m}_i \in \mathbb{R}^F$.
Given observations $\mathcal{C}$, our
goal is to find an estimate $\hat{\mathbf{H}}$ of a  3D ATF  $\mathbf{H}$ in a  target region.

Conventional learning-based approaches typically formulate sound field magnitude estimation as a regression problem, directly mapping sparse microphone inputs to dense field predictions. 
In this work, we instead adopt a generative perspective, aiming to model the distribution of 3D ATF magnitudes and guide the generation process using sparse microphone measurements. This formulation naturally aligns with FM, which provides an efficient generative framework for learning and sampling from complex data distributions.

\vspace{-4pt}
\subsection{Flow Matching for Sound Field Magnitude Estimation}
\vspace{-4pt}

\begin{figure}[t]
  \centering
  \includegraphics[width=\linewidth]{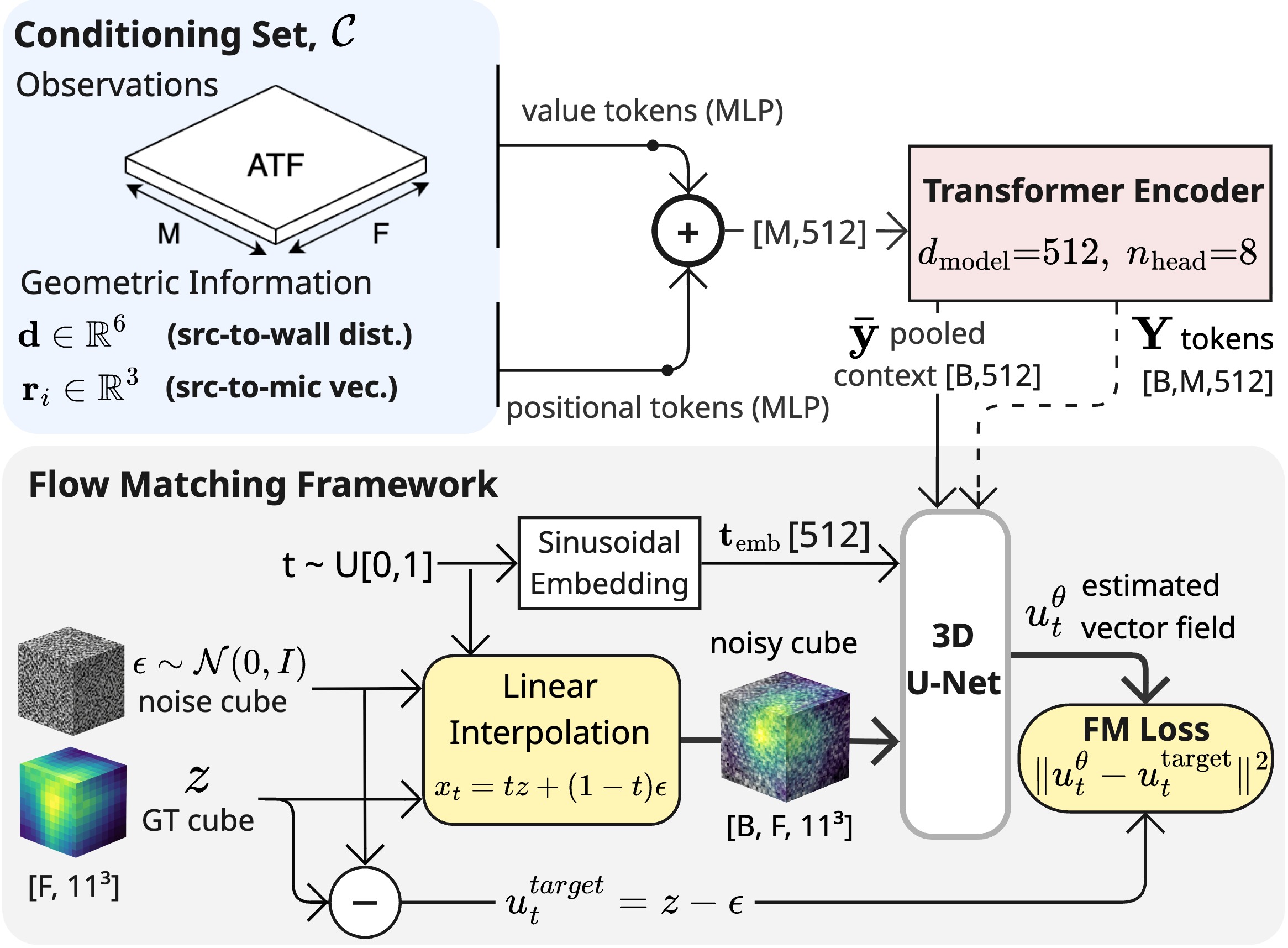}
  \vspace{-2em}
  \caption{Architecture diagram of SF-Flow. $B$ denotes the batch size.}
  \label{fig:diag}
\end{figure}

We formulate 3D ATF magnitude reconstruction as a conditional generation problem addressed with FM, following \cite{MIT25} for notation and definitions; Fig.~\ref{fig:diag} gives an overview of the framework. FM learns to transform samples from a simple prior distribution, $p_{\text{init}}$, into samples from the target data distribution $p_{\text{data}}$, where $p_{\text{data}}$ denotes the distribution of simulated 3D ATF magnitude cubes $\mathbf{H} \in \mathbb{R}^{F \times D \times H \times W}$ for each source position. The prior distribution $p_{\text{init}}$ is a standard Gaussian distribution of the same dimensionality. 


This transformation is modeled as a continuous-time process where the intermediate sample $\mathbf{X}_t$ evolves from $t=0$ to $t=1$. Here, $\mathbf{X}_0 \sim p_{\text{init}}$ is the initial random Gaussian noise cube and $\mathbf{X}_1 \approx \mathbf{H}$ is the terminal point of the trajectory, a physically plausible ATF magnitude cube from $p_{\text{data}}$. The dynamics of this evolution are governed by an Ordinary Differential Equation (ODE) defined by a time-dependent vector field $u_t$:
\begin{equation} 
\label{eq:ode}
\frac{d}{dt}\mathbf{X}_t = u_t(\mathbf{X}_t) \quad , \quad \mathbf{X}_0 \sim p_{\text{init}}~.
\end{equation}

The solution of this ODE defines a \textit{flow}, a function that maps noise samples to data, i.e., $\mathbf{X}_t =\psi_t(\mathbf{X}_0)$.
%
Intuitively, the flow induced by $u_t$ describes how the entire noise space is transported towards the data manifold of valid sound fields.
Thus, the problem of sound field magnitude estimation reduces to training a neural network ${u}_t^\theta(\cdot|\mathcal{C})$ that parametrizes this vector field, conditioned on the sparse set of microphone measurements $\mathcal{C}$.

The ideal training objective would be to minimize the mean squared error between the network's predicted vector field $u_{t}^{\theta}(x)$ and the true  marginal vector field $u_{t}^{\text{target}}(x)$.
However, the marginal vector field is intractable to compute because it requires integrating over the entire unknown data distribution. 
FM addresses this issue by replacing the marginal objective with a regression against the tractable conditional vector field $u_{t}^{\text{target}}(x|z)$, where $z \sim p_{\text{data}}$ denotes a single ground-truth ATF cube from the training set. Minimizing this Conditional FM (CFM) loss has been proven to be equivalent to minimizing the intractable marginal loss due to shared gradients \cite{lipman2023flow}:
\begin{equation}
 \mathcal{L}_{\text{CFM}}(\theta) = \mathbb{E}_{t \sim \text{Unif}, z \sim p_{\text{data}}, x \sim p_t(\cdot|z)} [\|u_{t}^{\theta}(x) - u_{t}^{\text{target}}(x|z)\|^2], 
\label{lcfm}
\end{equation}
where the conditional Gaussian vector field is given by \cite{MIT25}:
\begin{equation}
u_t^{\text{target}}(x|z) = \left( \dot{\alpha}_t - \frac{\dot{\beta}_t}{\beta_t} \alpha_t \right) z + \frac{\dot{\beta}_t}{\beta_t} x~.
\label{eq:cond_vect_field}
\end{equation}
A particularly effective choice is the linear Gaussian Optimal Transport (OT) path~\cite{MetaFM} with $\alpha_t=t$ and $\beta_t=1-t$, 
which linearly interpolates a data sample $z \sim p_{\text{data}}$ and a noise sample $\epsilon \sim \mathcal{N}(0, I)$ as $x_t = \alpha_t z + \beta_t \epsilon$.
Substituting this path into \eqref{eq:cond_vect_field} yields a constant conditional vector field that simplifies the training objective:
\begin{equation}
u_t^{\text{target}}(x_t|z) = z - \epsilon.
\end{equation}
In this formulation, unlike diffusion-based models that learn a complex time-dependent vector field, the network only needs to regress the difference between a ground-truth ATF cube and its corresponding noise sample.
Accordingly, the CFM loss \eqref{lcfm} reduces to:
\begin{equation}
\label{lcfm_ot}
\mathcal{L}_{\text{OT-CFM}}(\theta) = 
\mathbb{E}_{t,z,\epsilon} \big[\|u_{t}^{\theta}(x_t|\mathcal{C}) - (z-\epsilon)\|^2\big].
\end{equation}
This simplified and stable objective is a key factor behind the framework's fast convergence and its ability to generate high quality samples with only a few inference steps. Note that the FM framework specifies only the training objective and the target vector field, while imposing no restrictions on the architecture of the neural network used to predict $u_{t}^{\theta}$.
Our network architecture is given in \Cref{sec:gen_model}. At inference, the predicted ATF cube $\hat{\mathbf{H}}$ is obtained by integrating \eqref{eq:ode} forward from $\mathbf{X}_0 \sim \mathcal{N}(0,I)$ using $N=10$ Euler steps.

\begin{algorithm}[t]
\caption{Training procedure for SF-Flow}
\label{pseudo}
\begin{algorithmic}[1]
\Require Dataset $\mathcal{D}$ of ATF cubes $z \sim p_{\text{data}}$, model $u_t^\theta$
\For{each mini-batch}
  \State Sample a ground-truth ATF cube $z \sim \mathcal{D}$
  \State Sample a random time $t \sim \mathrm{Unif}(0,1)$
  \State Sample noise $\epsilon \sim \mathcal{N}(0, I)$
  \State Set $x_t = t z + (1 - t)\epsilon$ \hfill (Gaussian OT path)
  \State Sample observation count $M \sim \mathrm{Unif}\{5, 10, 20, 50\}$
  \State Form sparse measurement set $\mathcal{C}$ by randomly selecting $M$ positions and their corresponding magnitudes from $z$
  \State Compute loss: $\mathcal{L}(\theta) = \|u_t^\theta(x_t, \mathcal{C}) - (z - \epsilon)\|_2^2$
  \State Update $\theta$ via gradient descent
\EndFor
\end{algorithmic}
\end{algorithm}

\vspace{-4pt}
\subsection{Conditioning Encoder}
\label{sec:enc}
\vspace{-4pt}

To condition the generative process on the unordered, variable-sized set $\mathcal{C}$,
we employ a permutation-invariant set encoder based on the Transformer architecture. Each observation $(\mathbf{g}_i, \mathbf{m}_i)$ consists of a 9-dimensional coordinate
 vector $\mathbf{g}_i \in \mathbb{R}^9$,
 and its $F$-dimensional ATF magnitude vector $\mathbf{m}_i \in \mathbb{R}^F$. The coordinate and magnitude vectors are independently projected into $d_{\text{model}}=512$-dimensional embeddings via separate two-layer MLPs and summed
 to form the input token for each observation.
A three-layer Transformer encoder with $n_{\text{head}}=8$ attention heads then
refines these tokens through self-attention, capturing interactions among observations
while remaining permutation-invariant.
To accommodate variable set sizes, padding with a learned null token and binary masking
is applied.
The encoder produces two outputs: a sequence of per-observation tokens
$\mathbf{Y} \in \mathbb{R}^{M \times d_{\text{model}}}$ used for cross-attention in the
U-Net, and a global pooled context vector $\bar{\mathbf{y}} \in \mathbb{R}^{d_{\text{model}}}$
obtained by a masked mean over valid tokens, used for residual conditioning. Additionally, a per-frequency FiLM layer~\cite{perez2018film} applies
frequency-specific affine modulation to the U-Net input, using per-frequency
context vectors derived from the observation set via a lightweight MLP,
aiming for finer-grained spectral conditioning beyond the global pooled context.

\vspace{-4pt}
\subsection{Neural Network Architecture} \label{sec:gen_model}
\vspace{-4pt}

The generative model is a 3D U-Net that maps a noisy input cube
$\mathbf{X}_t \in \mathbb{R}^{F \times 11 \times 11 \times 11}$ to a predicted
vector field of identical shape. The network is conditioned on a sinusoidal time
embedding $\mathbf{t}_{\text{emb}} \in \mathbb{R}^{d_{\text{model}}}$ and both
outputs of the set encoder.
The encoder consists of two residual stages (channel sizes ${256, 512}$),   
each containing a residual block and a cross-attention block, followed by a 
bottleneck at $1024$ channels.   
In the residual block, the time embedding and global pooled context
$\bar{\mathbf{y}}$ are injected as additive biases after group normalization,
providing continuous global awareness of the observation set throughout all
encoder, bottleneck, and decoder stages.
The cross-attention block lets the 3D spatial features attend to the
per-observation token sequence $\mathbf{Y}$, anchoring local features to
specific microphone measurements.
The decoder mirrors the encoder with two up-sampling stages via transposed
convolutions, with skip connections concatenated from the corresponding encoder
stages. A $1{\times}1{\times}1$ convolution kernel maps back to $F$ output channels.
Reflective padding to $16^3$ is applied at the input and cropped at the     
output to accommodate the $2^2$ down-sampling factor.  

\vspace{-6pt}
\section{Experiments}
\label{sec:eval}
\vspace{-6pt}

\subsection{Experimental Setup} \label{sec:exp_setup}
\vspace{-4pt}

We simulated RIRs using the \texttt{pyroomacoustics} library~\cite{PRA_Scheibler_2018} for a room of dimensions \SI{4}{\meter}$\times$\SI{6}{\meter}$\times$\SI{3}{\meter}, with a  reverberation time (T60) of \SI{0.2}{\second}.
In each simulation, a sound source was placed at a position drawn uniformly over the entire room volume, excluding the target region.
The ground-truth sound field was represented by a 3D ATF magnitude cube, sampled at 1331 microphone positions on a uniform $11\!\times\!11\times\!11$ grid (\SI{0.1}{\meter} spacing) in a \SI{1}{\meter^3} target region centered in the room, i.e., \SI{1.5}{\meter}, \SI{2.5}{\meter}, and \SI{1.0}{\meter} away from the walls along the three room dimensions.
RIRs were simulated at a sampling rate of \SI{2000}{\hertz}, truncated to 128 samples, and converted to ATF magnitudes via the Fourier transform, yielding 64 frequency bins up to
\SI{1000}{\hertz}. 
The dataset, referred to as R1, comprised 1024 source positions, which were split into 820/102/102 for training, validation, and testing.
During training, the number of observations $M$ is drawn uniformly
from $\{5, 10, 20, 50\}$ per batch, and the $M$ microphone positions are sampled
without replacement from the full 1331-position grid.
The conditioning set $\mathcal{C}\!=\!\{(\mathbf{g}_i,\mathbf{m}_i)\}_{i=1}^M$
is padded to $M_{\max}=50$ with masking. The training procedure is shown in Algorithm~\ref{pseudo}.
For validation, each source is assigned a fixed, randomly predetermined set of $M_{\text{val}}=5$ observations, where $M_{\text{val}}$ denotes the number of conditioning observations available at validation time, ensuring reproducible and bias-free model selection. Every 200 iterations, full ATF predictions $\hat{\mathbf{H}}$ are generated via Euler ODE integration over all 102 validation sources, and the checkpoint with the lowest Log-Spectral Distortion (LSD, see Sec.~\ref{results}) is saved.
Experiments were conducted on an NVIDIA RTX A5000 GPU with a batch size of 4. The learning rate was linearly warmed up from $10^{-6}$ to $10^{-4}$ over the first 5000 iterations, followed by a cosine decay to $10^{-5}$ until iteration $10^{5}$, after which it was held constant.

We compare against the \textit{AE} baseline~\cite{Koyama:ForumAcusticum2025},
a conditioned autoencoder trained with the same discrete observation counts and identical experimental setup, using log-spectral distortion (LSD) error directly as the training loss, with the same
fixed-permutation validation protocol and $M_{\text{val}}\!=\!5$.
AE was trained for 1400 epochs, with a batch size of 1, using a step-wise learning rate schedule with decays from $10^{-3}$ to $10^{-4}$ and $10^{-5}$ at epochs 800 and 1200, respectively.
%
%
We also compare against \textit{KRR}~\cite{Koyama:ForumAcusticum2025}, a training-free Gaussian kernel ridge regression that estimates target magnitudes from observations (precision $10^{-2}$, regularization $10^{-3}$).
Source code and the dataset are available at \url{https://egerdem.github.io/sf-flow/}.

\begin{table}[t]
\centering
\caption{LSD, training time until best epoch, and per epoch times for full 3D ATF reconstruction from $M=5$ observations,
averaged over 102 test sources, across four frequency-bin ranges.}
\label{tab:lsd_r1}
\setlength{\tabcolsep}{2.7pt}
\begin{tabular}{p{1.1cm}cccc}
\toprule
\textbf{Method} & \textbf{0--20} & \textbf{0--30} & \textbf{0--40} & \textbf{0--64} \\
& \scriptsize{(312 Hz)} & \scriptsize{(468 Hz)} & \scriptsize{(625 Hz)} & \scriptsize{(1000 Hz)} \\
\midrule
\multicolumn{5}{l}{\small\textit{Dataset R1 (820 training sources) - All Methods}} \\[-2pt]                               
  \midrule
KRR            & $6.59 \pm 1.48$    & $8.10 \pm 1.35$    & $9.05 \pm 1.14$     & $10.67 \pm 1.11$    \\
\midrule
AE     & 
\makecell{$2.69 \pm 0.63$ \\ \SI{19.3}{\hour} / \SI{87}{\second}} & 
\makecell{$3.71\pm0.58 $ \\ \SI{21}{\hour} / \SI{94}{\second} } & 
\makecell{$\textbf{4.06$\pm$0.43}$ \\ \SI{21.4}{\hour} / \SI{96}{\second}} & \makecell{$\textbf{4.55±0.41}$ \\ \SI{24}{\hour} / \SI{108}{\second}} \\
\midrule
SF-Flow   & 
\makecell{\textbf{1.75$\pm$ 0.58} \\ \SI{5.8}{\hour} / \SI{20}{\second}} & 
\makecell{$\textbf{3.17$\pm$0.67}$ \\ \SI{4.8}{\hour} / \SI{20}{\second}} & 
\makecell{$4.16\pm0.63$ \\ \SI{3.6}{\hour} / \SI{20}{\second}} & 
\makecell{$5.56\pm0.52$ \\ \SI{2.4}{\hour} / \SI{20}{\second}} \\
\midrule
\multicolumn{5}{l}{\small\textit{Dataset R2 (13h,53s), R3 (18h,108s), R3Long (30h) - SF-Flow$^*$}} \\[-2pt]
\midrule
R2 &                                              
\makecell{$0.78\pm0.19$} &
\makecell{$1.57\pm0.33$} &                                                
\makecell{$2.55\pm0.43$} &                                                
\makecell{$4.44\pm0.41$} \\                                               
R3 &
\makecell{$0.66\pm0.16$} & 
\makecell{$1.25\pm0.21$} & 
\makecell{$2.25\pm0.45$} &
\makecell{$4.08\pm0.36$} \\
R3 Long &                                              
\makecell{$0.55\pm0.13$} &
\makecell{$0.97\pm0.19$} &                                                
\makecell{$1.66\pm0.30$} &                                                
\makecell{$3.67\pm0.36$} \\  
\bottomrule
\multicolumn{5}{l}{\scriptsize $^*$Trained for 900 (R2), 600 (R3), and 1000 (R3 Long) epochs.} \\
\end{tabular}
\end{table}

\begin{figure*}[t]
  \centering
  \includegraphics[width=0.85\textwidth,clip, trim=0 0.1cm 0cm 0cm]{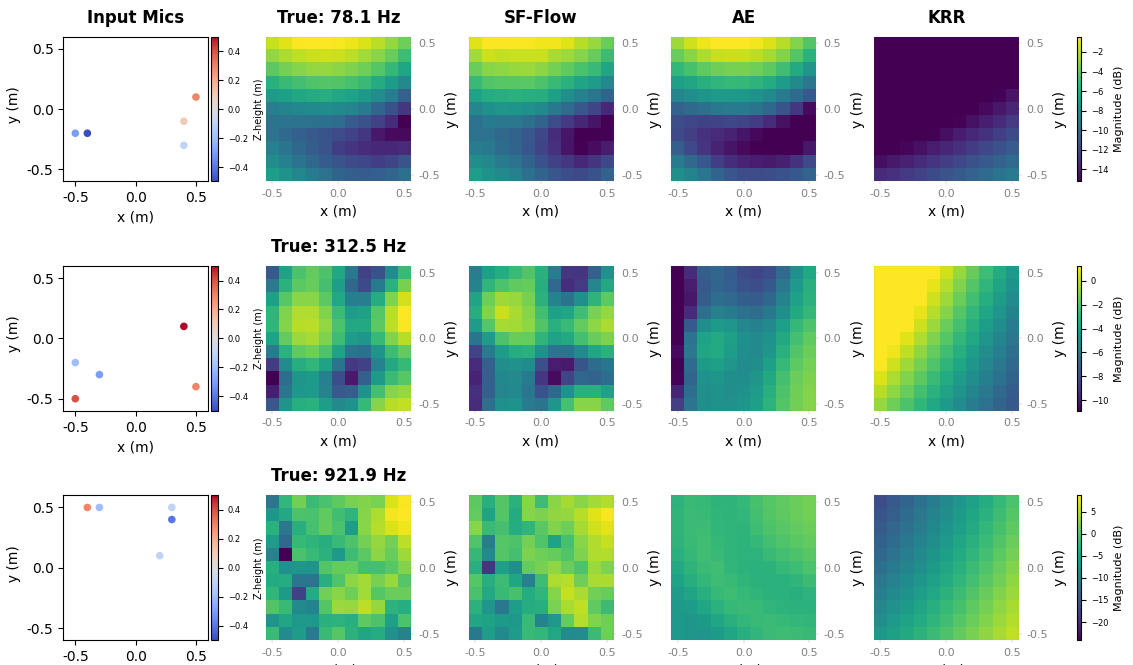}
  \caption{Estimated ATF magnitudes across three horizontal slices using the R1, 0--64 bins model. Rows represent different test sources and random sparse microphone configurations at heights (relative to the target-region center) of $-0.1$m ($78$Hz), $-0.2$m ($312$Hz), and $-0.3$m ($921$Hz). Columns compare the ground truth against SF-Flow, AE, and KRR.}
  \label{fig:sf_slices}
\end{figure*}


\vspace{-6pt}
\subsection{Results}
\label{results}
\vspace{-4pt}

\Cref{tab:lsd_r1} reports LSD for 3D ATF reconstruction from $M\!=\!5$ observations across
four frequency ranges, with $N\!=\!1331$ target positions, where $\text{LSD}\!=\!\tfrac{1}{N}\sum_{j=1}^{N}\
\sqrt{\tfrac{1}{F}\!\sum_{f=1}^F(\!\hat{H}\!_{f,j}-\!H\!_{f,j})^2}$, averaged over 102 test sources.
Both SF-Flow and AE  outperform KRR across all frequency ranges.
SF-Flow achieves lower LSD than AE up to 0--30 (468~Hz), despite AE being trained directly with LSD 
as its loss function. At higher frequencies, AE's direct LSD supervision becomes advantageous.
Wilcoxon signed-rank tests confirm these LSD differences are significant: SF-Flow wins on 101 and 97 of 102 sources for the 0--20 and 0--30 ranges ($p\!<\!10^{-17}$), whereas AE prevails for 0--40 ($p\!<\!0.005$) and 0--64 ($p\!<\!10^{-17}$). Zero-mean normalized cross-correlation (NCC)~\cite{damiano2025sound} shows SF-Flow yields higher spatial similarity for the 0--20, 0--30, and 0--40 ranges (0.92/0.77/0.61 vs.\ 0.81/0.65/0.57, $p < 10^{-7}$, where $1$ represents perfect similarity). This exposes AE's 0--40 LSD advantage as spatial over-smoothing. Conversely, AE's higher 0--64 NCC (0.47 vs.\ 0.38) reflects that point-wise metrics heavily penalize spatially displaced detail. Notably, per-epoch training time for AE grows from \SI{87}{\second} to \SI{108}{\second} as the frequency range widens (Table~\ref{tab:lsd_r1}). In contrast, SF-Flow requires a constant \SI{20}{\second} across all frequency ranges, taking a fraction of the overall time to reach its best checkpoint.

Fig.~\ref{fig:sf_slices} visualizes reconstructed ATF magnitudes at three horizontal          
slices for three test sources, using $M\!=\!5$  observations, for the
least accurate SF-Flow model trained on the full frequency range (0--64 bins).
At $f\!=\!78$\,Hz, both SF-Flow and AE accurately recover the large-scale field structure; SF-Flow additionally preserves finer spatial detail in the low-magnitude region. From $f\!=\!312$\,Hz, AE estimates become noticeably overly smooth, missing spatial variation that SF-Flow still partially retains. This tendency intensifies at $f\!=\!921$\,Hz, where SF-Flow retains realistic spatial variability despite its highest LSD error, in contrast to AE's 
near-uniform field estimation. This spatial smoothness may partly explain AE's competitive LSD at higher frequencies in Table~\ref{tab:lsd_r1}: regression toward the spatial mean reduces average error while losing reconstruction detail. KRR exhibits poor estimation across all scenarios.
Table~\ref{tab:lsd_m_ablation} reports LSD versus number of test observations $M$, demonstrating that SF-Flow enables full 3D field estimation from as few as a single observation, with                  
performance largely saturating beyond $M\!=\!5$. 


\textbf{Dataset Size Ablation.} The computational efficiency of SF-Flow allowed us to assess the effect of
training set size. We augmented the original 1024-source dataset (R1) with             
additional simulations to create R2 (4096 sources) and R3 (8192 sources),
keeping validation and test sets identical and the learning rate the             
same as R1. 
In the R1 setting, FM validation loss began to increase after an initial phase            
while validation LSD continued to drop slightly before plateauing, necessitating LSD-based checkpoint selection. Tuning hyperparameters, classifier-free guidance, or extended training did not improve upon this plateau.
For R2 and R3, FM validation loss correlates strongly with validation LSD, enabling
checkpoint selection by FM loss alone and eliminating the costly periodic full-inference  
validation LSD evaluation that accounted for nearly half of wall-clock time in R1.
Expanding to R2 and R3 substantially reduces LSD while maintaining training efficiency (Table~\ref{tab:lsd_r1}). Crucially, neither R2 (900 epochs), R3 (600 epochs),
nor R3 Long (R3 trained for 1000 epochs) had converged at their reported checkpoints, suggesting further gains are possible with continued training.

\begin{table}[t]
\centering
\caption{LSD vs.\ number of test observations $M$ (0--20 bins model)}
\label{tab:lsd_m_ablation}
\setlength{\tabcolsep}{3pt}
\begin{tabular}{l|ccccc}

\toprule

\textbf{$M$} & 1 & 5 & 10 & 20 & 50 \\ \hline
SF-Flow & $1.99$ & $1.75$ & $1.73$  &  $1.71$ & $1.72$ \\
AE  & 2.71 & 2.69 & 2.67 & 2.67 & 2.66 \\
\bottomrule
\end{tabular}
\end{table}



\vspace{-6pt}
\section{Conclusion and Future Work}
\label{sec:conc}
\vspace{-6pt}

We proposed SF-Flow, a method for estimating 3D ATF magnitudes from spatially sparse measurements based on FM. Our architecture using a 3D U-Net conditioned by a permutation-invariant set encoder enables reconstruction from an arbitrary number of measurements. Experimental results demonstrated that SF-Flow achieves performance competitive with AE while requiring substantially shorter training time, and retains spatial variability at higher frequencies where AE degrades toward overly smooth, near-uniform estimates. As future work, we plan to extend evaluation to multiple rooms and real recordings, and jointly model magnitude and phase to support downstream tasks such as 6-DoF audio rendering. Our results also motivate the development of shift-invariant evaluation metrics, since neither LSD nor NCC fully reflects reconstruction fidelity. Applying FM in a learned latent space, as commonly done in state-of-the-art applications, may offer further gains in accuracy and efficiency.




\vfill\pagebreak

\bibliographystyle{IEEEbib_mod}
\bibliography{strings,refs, str_def_abrv, skoyamalab_en}

\end{document}